\documentstyle [preprint,revtex]{aps}

\begin{document}
\begin{flushright}
IFP-457-UNC\\
TAR-029-UNC\\
VAND-TH-92-12\\
October 1992\\
Revised March 1993\\
\end{flushright}
\begin{center}
\Large
Positivity of Entropy in the Semi-Classical Theory of
Black Holes and Radiation
\end{center}
\normalsize
\begin{center}
{\large David Hochberg} and {\large Thomas W. Kephart}\\
{\sl Department of Physics and Astronomy, Vanderbilt University\\
Nashville, TN 37235}\\
and\\
{\large James W. York, Jr.}\\
{\sl Institute of Field Physics and\\
Theoretical Astrophysics and Relativity Group\\
Department of Physics and Astronomy\\
University of North Carolina, Chapel Hill, NC 27599-3255}\\
\end{center}
\vfill\eject
\begin{center}
ABSTRACT
\end{center}
\begin{abstract}
Quantum stress-energy tensors of
fields renormalized on a Schwarzschild background
violate the classical energy conditions near the black hole.
Nevertheless, the associated
equilibrium thermodynamical entropy
$\Delta S$ by which such fields augment the usual black hole
entropy is found to be positive. More precisely, the derivative of
$\Delta S$ with respect to radius, at fixed black hole mass, is
found to vanish at the horizon for {\it all} regular renormalized
stress-energy quantum tensors.
For the cases of conformal scalar fields and U(1) gauge fields, the
corresponding second derivative is positive, indicating that $\Delta S$
has a local minimum there. Explicit calculation shows that indeed
$\Delta S$ increases monotonically for increasing radius and is
positive. (The same conclusions hold for a massless spin 1/2
field, but the accuracy of the stress-energy tensor we employ
has not been confirmed, in contrast to the scalar and vector
cases).
None of these results would hold
if the back-reaction of the radiation on the spacetime geometry
were ignored; consequently, one must regard $\Delta S$ as arising
from both the radiation fields and their effects on the
gravitational field.
The back-reaction, no matter how ``small",
is therefore always significant in
describing thermal properties of the spacetime geometries and
fields near black holes.
\end{abstract}
\vfill\eject
\noindent
{\bf I Introduction}

A black hole can exist in thermodynamical equilibrium provided that
it is surrounded by radiation with a suitable distribution of
stress-energy. In the semi-classical approach, such radiation is
characterized by the expectation value of a stress-energy tensor
obtained by renormalization of a quantum field on the classical
spacetime geometry of a black hole. One can use such a stress-energy
tensor as a source in the semi-classical Einstein equation,
$$G^{\mu}_{\nu} = 8\pi <T^{\mu}_{\nu}>_{renormalized},\eqno(1)$$
to calculate the change effected by the stress-energy tensor
in the black hole's spacetime metric. This is the ``back-reaction"
problem associated with the spacetime geometry of a black hole in
equilibrium.

In this paper we use solutions of back-reaction problems of the
above type to compute the thermodynamical entropy $\Delta S$ by
which quantum fields augment the usual Bekenstein-Hawking black
hole entropy $S_{BH} = (1/4) A_H {\hbar}^{-1}$, where $A_H$ is
the area of the event horizon (Units are chosen such that
$G = c = k_{B} = 1$, but ${\hbar} \neq 1$.).
We consider explicitly the case of a Schwarzschild black hole
surrounded by either a massless conformal scalar field or a U(1)
gauge field (Maxwell field). (A massless spin 1/2 field is treated
in the Appendix, but the accuracy of its stress-energy tensor
has not to our knowledge been checked, in contrast to the
conformal scalar and vector fields.) We show in all these cases
that $\Delta S$ is positive.

Our investigation shows rigorously that for {\it all} possible
regular stress-energy tensors, the radial derivative of $\Delta S$
vanishes at the horizon, for fixed black-hole mass; that is, $\Delta S$
has there a local extremum with respect to radius. The form of the
second derivative gives the criterion for a local minimum, which indeed
occurs in all cases we have considered. Then by explicit calculation
we show that $\Delta S$ is positive and
monotonically increasing for increasing
radius. Therefore the local minimum of $\Delta S$ at the
horizon is
the only one and is its global minimum.
As a consequence, the entropy is amenable to statistical interpretation.
None of these features holds if the back-reaction of the fields
on the spacetime metric is ignored. In this sense, $\Delta S$
must be regarded as arising from both the quantized radiation
fields {\it and} from their effects on the gravitational
field.

We shall see, from the properties of the renormalized stress-energy
tensors we employ and of the semi-classical Einstein equation, that
we can obtain
accurate fractional corrections to the metric
only in $O(\epsilon)$, where $\epsilon = {\hbar} M^{-2}$,
$M_{Pl} = {\hbar}^{1/2}$ is the Planck mass
and $M$ is the mass of the black hole. Because the usual black hole
entropy
$S_{BH} = (4\pi M^2) {\hbar}^{-1} = O(\epsilon^{-1})$, corrections
to $S_{BH}$ can be obtained in $O(\epsilon^0) = O(1)$ from
fractional corrections of $O(\epsilon)$ in the metric. It turns
out that these corrections are of the same order as the naive
flat space radiation entropy $(4/3)a\, T^3_{H} V$, where
$a = ({\pi}^2/{15 \hbar^3})$, $T_{H} = \hbar (8\pi M)^{-1}$ is
the uncorrected Hawking temperature of a Schwarzschild black hole, and
$V$ is the flat space volume. From this fact alone it follows that the
back-reaction cannot be ignored.\\

\noindent
{\bf II Stress-Energy Tensors}

Stress-energy tensors renormalized on a Schwarzschild background have
been obtained in exact form for conformal scalar fields and for U(1)
gauge fields, respectively, by Howard [1] and
by Jensen and Ottewill [2]. Both results can be written in the form
$$<T^{\mu}_{\nu}>_{renormalized} = <T^{\mu}_{\nu}>_{analytic} +
 \left( {{\hbar} \over {\pi^2 (4M)^4}} \right)
\Delta^{\mu}_{\nu},\eqno(2)$$
where the analytic piece, in the case of a conformal scalar field, was
given by Page [3]. The term $\Delta^{\mu}_{\nu}$ is obtained from
a numerical evaluation of a mode sum. The numerical piece is small
compared to the analytic piece, and we do not include it in the
calculations in this paper. This does
not change any of our results qualitatively because both pieces
separately obey the required regularity and consistency conditions.
The analytic piece has the exact trace anomaly in both cases.

The stress-energy tensors satisfy
${\hat \nabla}_{\mu} <T^{\mu}_{\nu}> = 0$
on the Schwarzschild background with metric
$${\hat g}_{\mu \nu} = {\rm diag} \left[ -(1 - {{2M} \over r}),
(1 - {{2M} \over r})^{-1}, r^2, r^2 \sin^2 \theta
\right] .\eqno(3)$$
These tensors represent the stress-energy distribution required to
equilibrate the black hole with its own Hawking radiation. Each
satisfies $<T^t_t>=<T^r_r>$ at the horizon $r = 2M$, which is
required for regularity of the spacetime geometry [3].
Each has the asymptotic form of a flat spacetime radiation
stress-energy tensor at the uncorrected Hawking temperature at
infinity of an ordinary Schwarzschild black hole, denoted here by
$T_{H} = \hbar (8\pi M)^{-1}$.

Dropping the angular brackets and displaying the analytic piece, one has
for the conformal scalar field [3]
$$T^t_t = -{1 \over 3}aT_H^4 ({1 \over 2}) \left(  3 + 6w + 9w^2 + 12w^3
 + 15w^4 + 18w^5 - 99w^6 \right) ,\eqno(4)$$
$$T^r_r = {1 \over 3}aT_H^4 ({1 \over 2}) \left( 1 + 2w + 3w^2 + 4w^3
 + 5w^4 + 6w^5 + 15w^6 \right) ,\eqno(5)$$
$$T^{\theta}_{\theta} = T^{\phi}_{\phi} = {1 \over 3}aT_H^4 ({1 \over 2})
\left( 1 + 2w + 3w^2 + 4w^3 + 5w^4 + 6w^5 - 9w^6 \right) ,\eqno(6)$$
where $w \equiv 2M/r$. We have displayed the factor $(1/2)$ explicitly
because the scalar field has one helicity state while the vector field
below has two. It is convenient in what follows to write
$${1 \over 3}a T_H^4 = {{\epsilon} \over {48 \pi K M^2}},\eqno(7)$$
where $K = 3840 \pi$. For the U(1) vector field, we have [2]
$$T^t_t = -{1 \over 3}aT_H^4 \left( 3 + 6w + 9w^2 + 12 w^3 - 315w^4
+ 78w^5 - 249w^6 \right),\eqno(8)$$
$$T^r_r = {1 \over 3}aT_H^4 \left( 1 + 2w + 3w^2 - 76w^3 + 295w^4 - 54w^5
 + 285w^6 \right),\eqno(9)$$
$$T^{\theta}_{\theta} = T^{\phi}_{\phi} = {1 \over 3}aT_H^4
\left( 1 + 2w + 3w^2 + 44w^3 - 305w^4 + 66w^5 - 579w^6 \right).\eqno(10)$$
In both cases $T^r_r > 0$ and the energy density $-T^t_t$ is negative
in the vicinity of the event horizon, thus violating the weak energy
condition. For the scalar field, the energy density is negative from
$r = 2M$ to $r \approx 2.34M$ and for the vector field from
$r = 2M$ to $r \approx 5.14M$. Both tensors also violate the
dominant energy condition in a region surrounding and bordering
on the horizon.\\

\noindent
{\bf III Back-reaction on the Metric}

We obtain fractional corrections $h^{\alpha}_{\nu}$ to the metric by
setting
$$g_{\mu \nu} = {\hat g}_{\alpha \mu} [\delta^{\alpha}_{\nu} +
 \epsilon\, h^{\alpha}_{\nu}]\eqno(11)$$
in the semi-classical Einstein equation (1). We work in linear order
in $\epsilon$ as required by ${\hat \nabla}_{\mu} T^{\mu}_{\nu} = 0$
and ${\hat \nabla}_{\mu} (\delta G^{\mu}_{\nu}) = 0$, where
$\delta G^{\mu}_{\nu}$
is the Einstein operator linearized on a background satisfying
${\hat G}^{\mu}_{\nu} = 0$.
The corrected geometry will be taken to be
static and spherically symmetric. Working out the equations as in [4],
we find the corrected metric can be written as
$$ds^2 = -\left(1 - {{2m(r)} \over r} \right)
 \left(1 + 2 \epsilon {\bar \rho}(r) \right) dt^2
 + \left(1 - {{2m(r)} \over r} \right)^{-1} \!dr^2 + r^2 d{\omega}^2,
\eqno(12)$$
where $d\omega^2$ is the standard metric of a normal round unit sphere.
To obtain $m(r)$ and ${\bar \rho}(r)$ requires only simple radial
integrals involving $T^t_t$ and $T^r_r$. The angular components enter
linearized Einstein equations that hold automatically by
virtue of ${\hat \nabla}_{\mu} T^{\mu}_{\nu} = 0$ in a static
spherical geometry.

The mass function $m(r)$ has the form
$$m(r) = M(1 + \epsilon\,\mu(r) + \epsilon\, C K^{-1}),\eqno(13)$$
with
$$\mu(r) = {1 \over {\epsilon M}}\, \int_{2M}^{r} (-T^t_t)\,
4\pi{\tilde r}^2 \, d{\tilde r},\eqno(14)$$
so $\mu(r)$ vanishes at the horizon. In (13), $C$ is an
undetermined integration constant that inspection of (12)
shows is to be absorbed into $M$ to obtain a renormalized mass
for the black hole. Thus, setting $g^{rr} = 0$ shows that
$r = 2m = 2M(1 + \epsilon\,C K^{-1}) = 2M_{renormalized}$
locates the event horizon.
Note that, to the order we are working, we can write
$m(r) = M(1 + \epsilon\,C K^{-1})(1 + \epsilon\, \mu(r)) \equiv
M_{ren}(1 + \epsilon\, \mu(r)).$
The renormalized mass will not be
distinguished notationally from the original Schwarzschild
mass $M$ in what follows, as the bare Schwarzschild mass has
no physical meaning in the back-reaction problem.
Therefore, we write
$$m(r) = M(1 + \epsilon\,\mu(r)) \equiv M + M_{rad}(r)\eqno(15)$$
where, using (14), we see that
$M_{rad} = \epsilon\,M\,\mu$ is the usual expression for the effective
mass of a spherical source.

For the scalar field, denoted where necessary by a subscript ``$s$",
one finds [4]
$$K\,\mu_s = {1 \over 2}({2 \over 3}w^{-3} + 2w^{-2}
+ 6w^{-1} - 8\ln(w) - 10w - 6w^2 + 22w^3 - {{44} \over 3}).\eqno(16)$$
For the vector field, denoted by a subscript ``$v$", one finds [5]
$$K\,\mu_v = {2 \over 3}w^{-3} + 2w^{-2} + 6w^{-1} - 8 \ln(w)
 + 210w - 26w^2 + {{166} \over 3}w^3 - 248.\eqno(17)$$
In both (16) and (17), we note that the first term on the right, multiplied
by $\epsilon\, M K^{-1}$, gives the naive flat-space value $a\, T_H^4 V$
for radiation energy.

The metric is completed by a determination of ${\bar \rho}$ which, like
$\mu$, can be found from an elementary integration. Defining
$$K\,{\bar \rho} \equiv K\,\rho + k,\eqno(18)$$
where $k$ is a constant of integration, we have
$$\rho = {1 \over {\epsilon}} \int_{2M}^{r} (T^r_r - T^t_t)
({\tilde r} - 2M)^{-1} 4\pi{\tilde r}^2 \, d{\tilde r}.\eqno(19)$$
For the scalar field, one finds [4] ($K {\bar \rho}_s = K \rho_s + k_s$)
$$K \rho_s = {1 \over 2} \left( {2 \over 3}w^{-2} + 4w^{-1} -
8\ln(w) - {{40} \over 3}w - 10w^2 - {{28} \over 3}w^3 + {{84} \over 3}
\right).\eqno(20)$$
Note that at the horizon $r = 2M$, or $w = 1$, we have
$\rho_s(1) = 0$. The constant $k$ for the scalar (vector)
is denoted
$k_s$ $(k_v)$ and will be determined
below by a boundary condition.
Similarly, for the vector field we have $K\, {\bar \rho}_v =
 K\,\rho_v + k_v$, where [5]
$$K\,\rho_v = {2 \over 3}w^{-2} + 4w^{-1} - 8\ln(w) +
{{40} \over 3}w + 10w^2 + 4w^3 - 32,\eqno(21)$$
and $\rho_v(1) = 0$ at $w = 1$.

Because both radiation stress-energy tensors are asymptotically
constant, it is clear that the system composed of black hole plus
equilibrium radiation must be put in a finite ``box". Otherwise, the
fractional corrections $\epsilon\,h^{\alpha}_{\nu}$ to the metric
would not remain small for sufficiently large radius. Physically, this
means that the radiation
in a box that is too large
would collapse onto
the black hole, producing a larger one. Hence, we must choose the radius
$r_o$ of the box such that it is less than the second positive root
$r_{\ast}$ for $r$ in $g^{rr} = 0$ (the first zero corresponds to the
horizon $r = 2M$).
We shall also assume that the box radius $r_o$ is sufficiently large
that the stress-energy tensors we employ, which were constructed for
infinite asymptotically flat spacetime, are a good approximation.
Clearly, a finite radius would cut out some of the radial modes that
were used in these calculations. However, if $r_o$ is somewhat greater
than the longest wavelength characteristic of Hawking radiation, which
in turn is associated with the least-damped quasi-normal mode of
lowest angular momentum for the field in question, then this effect
should be negligible. This wavelength $\lambda_*$ is about $42 M$
for the conformal scalar field and is smaller for the higher-spin
massless fields. Also, if $r_o > \lambda_*$, the explicit nature of
the walls of the box ({\it e.g.}, adiabatic {\it versus} diathermic)
should not be important. For these reasons we shall assume throughout
the remainder of this work that $\lambda_* < r_o < r_*$.
(Of course, one must also assume that $M \stackrel {>}{\sim} M_{Pl}$,
in any treatment
based on (1).) If the radius $r_o$ were to approach the horizon, then
explicit size and boundary effects would have to be taken into account
in the construction of $<T^{\mu}_{\nu}>$, as shown in the work
of Elster [6,7].

One convenient way to fix the constants $k_s$ and $k_v$ is to
impose a microcanonical boundary condition [4].
We fix $r_o$ and imagine placing there an ideal massless perfectly
reflecting wall. Outside $r_o$, we then have an ordinary
Schwarzschild spacetime
$$ds^2 = - \left( 1 - { {2m(r_o)} \over r} \right) dt^2 +
\left( 1 - { {2m(r_o)} \over r} \right)^{-1} \! dr^2 + r^2 d\omega^2,
\eqno(22)$$
for $r \geq r_o$.
Continuity of the three-metric induced by metrics (12) and (22)
on the world tube $r = r_o$ fixes the constant $k$, i.e., $k_s$
or $k_v$, in $\bar \rho$ by the relation
$$k = - K\, \rho(r_o).\eqno(23)$$
There are finite discontinuities in the extrinsic curvature
of the world tube
$r = r_o$ [4], but these, and other properties of the box
wall, are of no interest in the present analysis, as we
argued above.
The spacetime
geometry, including back-reaction, is now completely determined by
(22) for $r \geq r_o$, and for $r \leq r_o$ by
$$ds^2 = -\left( 1 - { {2m(r)} \over r} \right)
[1 + 2\epsilon\,(\rho(r) - \rho(r_o)) ] dt^2 +
\left( 1 - { {2m(r)} \over r} \right)^{-1} \! dr^2 + r^2 \,d\omega^2
.\eqno(24)$$\\

\vfill\eject
\noindent
{\bf IV Temperature}

If we release a small packet of energy from a closed
box containing a black hole through a long
thin radial tube, it will undergo a red-shift and approach the
asymptotic temperature
$$T_{\infty} = {{\kappa_H \,\hbar} \over {2\pi}},\eqno(25)$$
where $\kappa_H$ is the surface gravity of the event horizon.
For an ordinary Schwarzschild black hole (ignoring the radiation), one
finds $\kappa_H = (4M)^{-1}$
and $T_{\infty} = T_H = \hbar (8\pi M)^{-1}$.
However, the stress-energy of the radiation changes the surface gravity
of the horizon to
$$\kappa_H = {1 \over {4M}} \left[ 1 + \epsilon ({\bar \rho} - \mu) +
8\pi r^2\, T^t_t \right] |_{r=2M},\eqno(26)$$
as a straightforward calculation shows [4]. With the
microcanonical boundary conditions, we can use (23) to obtain from
(25) and (26)
$$T_{\infty} = {{\hbar} \over {8\pi M}} \left[ 1 - \epsilon\, \rho(r_o)
 + \epsilon\, n K^{-1} \right],\eqno(27)$$
where $n$ takes the value $n_s = 12$ for the scalar field and
$n_v = 304$ for the vector field. The local temperature at the
boundary of the box is obtained by blue-shifting (27) from infinity
back to $r_o$. We find from
$$T_{loc} = T_{\infty} [-g_{tt}(r_o)]^{-1/2},\eqno(28)$$
that
$$T_{loc}(r_o) = {{\hbar} \over {8\pi M}} \left[ 1 - \epsilon\, \rho(r_o)
 + \epsilon\,n K^{-1} \right] \left[ 1 - {{2m(r_o)} \over {r_o}}
\right]^{-1/2} .\eqno(29)$$
The temperature $T_{loc}$, unlike $T_{\infty}$, is actually
{\it independent}
of the boundary condition that determines
the constant $k$, as explained in detail
in [4].
Indeed, it can be readily verified by the reader that $k$ cancels
out in $O(\epsilon)$ in the expression (28) for $T_{loc}$.
{\it Either} measure of temperature, $T_{\infty}$ or $T_{loc}$, can be
used to calculate the same entropy in conjunction with an appropriate
measure of energy. This is quite important: it means that the
specific boundary condition chosen does {\it not} affect the
calculated entropy, as we shall see below.

\noindent
{\bf V Thermodynamical Entropy}

One way to calculate the entropy is as follows. Fix the radius $r_o$
of a closed box. The measure of energy in the box conjugate to the
asymptotic inverse temperature $\beta_{\infty} \equiv T^{-1}_{\infty}$
is then the Arnowitt-Deser-Misner (ADM) mass $m(r_o)$ determined at
spatial infinity. The first law of thermodynamics for slightly
differing equilibrium configurations tells us that
$$dS = \beta_{\infty}\, dm \qquad (dr_o = 0),\eqno(30)$$
where $S(r_o)$ is the total entropy in the box. By this method we seem
to obtain only the total entropy $S(r_o)$ rather than the distribution
of entropy in the given box, $S(r)$, for $r \leq r_o$, where $S(r)$
denotes the total entropy inside the radius $r$.
However, the
latter can be obtained by using the quasi-local energy $E$ [8-11], which
for static spherical metrics like those treated here is given for
any radius $r \leq r_o$ by
$$E(r) = r - r [g^{rr}(r)]^{1/2}, \eqno(31)$$
with $g^{rr}(r)$ determined by (24), the metric for $r \leq r_o$.
This energy, unlike $m$, does not depend on asymptotic flatness
in its definition, nor even on the existence of an asymptotically
flat region [10,11]. Furthermore, even the ``normalization" of the
zero of energy [10,11] that is incorporated in $E$ as given in (31)
does not affect the calculated entropy, as it certainly should not.
(This ``normalization" is intended to make $E$ approach the ADM mass
in an asymptotically flat region, if such a region exists.)
Similarly, the inverse local temperature
$\beta(r) \equiv T^{-1}_{loc}(r)$, $r \leq r_o$, is independent
of the boundary conditions as mentioned above.
Hence, {\it the value of the entropy depends neither on the zero
of energy nor on the existence of an asymptotic region.}

Therefore, to obtain $S(r)$, in place of (30) we can write
$$dS = \beta\, dE \qquad (dr = 0, \, r \leq r_o).\eqno(32)$$
Choosing $M$ and $r$ as independent variables, and fixing $r$, we
can readily integrate (32) to obtain $S$ up to a function of $r$
and a constant. From (29) we have
$$\beta (r) = {{8\pi M} \over {\hbar}} \left[1 + \epsilon\,
\rho (r) - \epsilon\, n K^{-1} \right] \left[1 - {{2m(r)} \over r}
\right]^{1/2}
,\eqno(33)$$
and from (15), (24), and (31), holding $r$ fixed,
$$dE = \left[ 1 - \epsilon \, \mu + \epsilon\, M
{{\partial \mu} \over {\partial M}} \right]^{-1/2} \,
\left[1 - {{2m(r)} \over r} \right]^{-1/2} \, dM. \eqno(34)$$
One can see directly for any $r \leq r_o$ that
$\beta_{\infty} dm = \beta\, dE$ where, of course, one replaces
$r_o$ by $r$ in the formulas for $\beta_{\infty}$ and $m$ to
establish this result.
This equality means that we can
calculate $S(r)$ for any $r \leq r_o$. The key point of this
discussion is that one can think
of adding layer upon layer of entropy, associated with the black
hole and a given $<T^{\mu}_{\nu}>$ that is valid from
$r = 2M$ to $r = r_o$, beginning at $r = 2M$ and ending at
$r = r_o$. (Additivity of entropy in configurations analogous
to this case is established in [12], but our method here establishes
it independently.)

Observe that from fractional changes of $O(\epsilon)$ in the
metric, which affect the surface gravity and temperature in this
order, we are able to calculate from (32) departures of
$O(\epsilon^0) = O(1)$ from the usual black hole entropy
$S_{BH} = (4\pi M^2) {\hbar}^{-1} = 4\pi \epsilon^{-1}$.
But in fact all of the corrections to the entropy are of the
{\it same} order as the naive flat-space entropy itself:
$${4 \over 3}a\, T^3_H V = {4 \over 3}({{\pi^2} \over {15 \hbar^3}})
({{\hbar} \over {8\pi M}})^3 ({4 \over 3}\pi r^3) =
{{8\pi} \over K} ({8 \over 9} w^{-3})
= O(1)\times w^{-3}.\eqno(35)$$

The $\hbar$'s in (35) cancel out, leaving only a function of
$w = 2M r^{-1}$.

Combining (33) and (34) yields
$$dS = {{8\pi M} \over {\hbar}} dM + 8\pi \left[ w^{-1}
(\rho - \mu) + {{\partial \mu} \over {\partial w}} -
n\, K^{-1} w^{-1} \right]\,dw,\eqno(36)$$
with $dr = 0$. Integration of (36) gives an expression
of the form
$$S = {{4\pi M^2} \over {\hbar}} + \Delta S(w) + f({{r} \over
 {\hbar^{1/2}} }), \qquad (1 \leq w \leq w_o = 2M/{r_o})
\eqno(37)$$
where the first term is the usual Bekenstein-Hawking expression
$S_{BH}$ for the black hole entropy, the second term is a function
of $w$ determined up to an additive integration
constant by the second term
on the right of (36), and $f$ is a dimensionless function of
$r$ that does not depend on $M$. The appearance of a function $f$
in (37) can be understood as follows. Since our problem involves
three mass or length scales $M_{Planck} = \hbar^{1/2}$, the mass
of the black hole, $M$, and a radius $r \leq r_o$, there are,
for a given $r$, exactly
three dimensionless parameters one can define, namely, $\epsilon =
\hbar M^{-2}$, $w = 2M/{r}$ and ${r}/{\hbar^{1/2}}$. However, the
first two terms on the right of (37) depend only on $\epsilon$ and
$w$, respectively. Thus, if the entropy $S$ depends on
${r}/{\hbar^{1/2}}$, it can only do so through a separate function
of this parameter.

Let us first dispose of the dimensionless function $f$, which clearly
can depend only on $(r/{\hbar}^{1/2})$, where $\hbar^{1/2}$ is
the Planck length in our units. It seems that such a term could only
arise in a theory taking quantum gravity into explicit account because
the semi-classical theory has incorporated the dimensionless terms
involving $\hbar/M^2$ and $2M/r$. (Of course, quantum gravity could
modify terms of these latter two types quantitatively.)
On dimensional grounds, therefore, we take $f = 0$
in the semi-classical theory. (A formal argument that $f = 0$ based
on [9] can be constructed [13].)
The possibility of an additive constant will be
discussed when we treat $\Delta S$ below.

In considering $\Delta S$, which will be given explicitly below, we
first note the significant property that
$${{\partial (\Delta S)} \over {\partial w}} = 8\pi
\left[ w^{-1} (\rho - \mu) + {{\partial \mu} \over {\partial w}}
- n\,K^{-1} w^{-1} \right]\eqno(38)$$
{\it vanishes at the horizon} $w =1$. Therefore, for a fixed black
hole mass $M$, the derivative with respect to $r$ of
$\Delta S$
vanishes at the horizon.
Thus $\Delta S$ has a local extremum with respect to $r$ at
the horizon.
This result
follows from several general features that will be enjoyed by
{\it all} regular renormalized stress-energy tensors on the
Schwarzschild background and the back-reactions they induce, not
just the cases analyzed here. First, $\mu$ vanishes at the
horizon by virtue of the black hole's mass having been suitably
renormalized. Second, $\rho$ vanishes at the horizon, as follows
from (19) and the regularity condition $T^t_t = T^r_r$ at the
horizon [3]. More precisely, we have that
$${\rm lim}_{w \rightarrow 1^{+}} \left(
{{T^t_t - T^r_r} \over {1 - w}} \right) \qquad {\rm exists}.
\eqno(39)$$
Third, the last two terms on the right of (38) add to zero at the
horizon because there the Hamiltonian
constraint $(G^t_t - 8\pi T^t_t = 0)$ holds.
Furthermore, note that if the fractional effects of $O(\epsilon)$
in the temperature induced by the back-reaction were neglected, the
derivative (38) would not vanish at the horizon, a property
that the reader can verify.

Is the local extremum of $\Delta S$ at the horizon a local minimum? To
answer this we calculate
$${{\partial^2 (\Delta S)} \over {\partial w^2}} = 8\pi \left[
-w^{-2} (\rho - \mu) + w^{-1}( {{\partial \rho} \over {\partial w}}
 - {{\partial \mu} \over {\partial w}}) +
{{\partial^2 \mu} \over {\partial w^2}} + n K^{-1} w^{-2} \right],
\eqno(40)$$
which becomes, at the horizon $w = 1$,
$${{\partial^2 (\Delta S)} \over {\partial w^2}}|_{w = 1} =
8\pi \left( {{\partial \rho} \over {\partial w}} +
{{\partial^2 \mu} \over {\partial w^2}} \right) |_{w = 1} \eqno(41)$$
or , equivalently, with $M$ fixed,
$${{\partial^2 (\Delta S)} \over {\partial r^2}}|_{r = 2M} =
{{32\pi^2 M^2} \over {\hbar}} \left[ 4M\, {{\partial (- T^t_t)} \over
{\partial r}} - 8 T^r_r - ({{T^r_r - T^t_t} \over {1 - 2M/r}}) \right]
|_{r = 2M}. \eqno(42)$$
Hence we need only examine the stress-tensors. In all the cases we
consider (conformal scalar, vector, massless fermion), (41) and (42)
are positive so that $\Delta S$ takes a local minimum with respect
to radius at the horizon. This suggests, but does not prove, that
$\Delta S$ is non-negative.

The local minimum of $\Delta S$ at the horizon and the fact that
$S_{BH}$ in the expression (37) for the total entropy $S$ contains
the {\it renormalized} mass $M$ of the hole motivate the choice of the
remaining additive constant in $\Delta S$, which can only be a pure
number, to be such that $\Delta S = 0$ at $w = 1$.
For $w = 1$, with no ``room" for the fields to contribute anything
further, one then obtains only the Bekenstein-Hawking entropy
$(1/4) A_{H} {\hbar}^{-1}$, as would be expected.
With the choice $\Delta S(w = 1) = 0$,
we obtain for the conformal scalar field [14,15]
$$\Delta S_s = {{8\pi} \over K}({1 \over 2}) \left( {8 \over 9}w^{-3}
 + {8 \over 3}w^{-2} + 8w^{-1} + {{32} \over 3} \ln(w)
 - {{40} \over 3}w - 8w^2 + {{104} \over 9}w^3 - {{16} \over 9}
\right)\eqno(43)$$ for $1 \geq w \geq w_o$.
Similarly, for the electromagnetic or U(1) gauge field we find
$$\Delta S_v = {{8\pi} \over K} \left( {8 \over 9}w^{-3}
 + {8 \over 3}w^{-2} + 8w^{-1} - 96\ln(w) +
{{40} \over 3}w - 8w^2 +{{344} \over 9}w^3 - {{496} \over 9}
\right).\eqno(44)$$
In both expressions, the naive flat-space radiation entropy term (35)
appears as the first term on the right. Both $\Delta S_s$ and
$\Delta S_v$ are positive for $1 \geq w \geq w_o > w_* = 2Mr_*^{-1}$
and vanish at
$w = 1$. Hence, in that they are positive, both are amenable
to arguments relating thermodynamical and statistical entropy.
It has not heretofore been evident that this desirable feature
would be present in the semi-classical theory.
The reader can verify, by omitting the back-reaction terms in the
inverse temperature (33), that not only is the vanishing slope
of $\Delta S$ at $w = 1$ lost, but also that the value of the
resulting ``$\Delta S$", normalized as above, is no longer
positive for the range $1 \geq w \geq w_o$.
In this fundamental sense,
we conclude that the back-reaction, however small
quantitatively in its effects on the metric near a black hole,
can never be regarded as negligible.\\

\noindent
{\bf Acknowledgments}

J.W.Y. thanks G.L. Comer for helpful discussions and correspondence.
This research was supported by DOE Grant No. DE-FGO5-85ER40226
(D.H. and T.W.K.) and by National Science Foundation grants
PHY-8407492 and PHY-8908741 (J.W.Y.).\\

\noindent
{\bf References}

\noindent
[1]. K.W. Howard, Phys. Rev. D{\bf 30}, 2532 (1984).

\noindent
[2]. B.P. Jensen and A. Ottewill, Phys. Rev. D{\bf 39}, 1130 (1989).

\noindent
[3]. D.N. Page, Phys. Rev. D{\bf 25}, 1499 (1982).

\noindent
[4]. J.W. York, Phys. Rev. D{\bf 31}, 775 (1985).

\noindent
[5]. D. Hochberg and T.W. Kephart, Phys. Rev. D{\bf 47}, 1465 (1993).

\noindent
[6]. T. Elster, J. Phys. A: Math. Gen. {\bf 16}, 989 (1983).

\noindent
[7]. T. Elster, Class. Quantum Grav. {\bf 1}, 43 (1984).

\noindent
[8]. J.W. York, Phys. Rev. D{\bf 33}, 2092 (1986).

\noindent
[9]. J.D. Brown, G.L. Comer, E.A. Martinez, J. Melmed, B.F. Whiting\\
and J.W. York, Class. Quantum Grav. {\bf 7}, 1433 (1990).

\noindent
[10]. J.D. Brown and J.W. York, Contemporary Mathematics, {\bf 132},
129 (1992).

\noindent
[11]. J.D. Brown and J.W. York, Phys. Rev. D{\bf 47}, 1407 (1993).

\noindent
[12]. E.A. Martinez and J.W. York, Phys. Rev. D{\bf 40}, 2124 (1989).

\noindent
[13]. G.L. Comer and J.W. York, {\sl Integrability of the entropy},
(unpublished, 1991).

\noindent
[14]. J.W. York, {\sl Entropy of a conformal scalar field and
a black hole}, (unpublished, 1985).

\noindent
[15]. G.L. Comer, {\sl The thermodynamic stability of systems
containing black holes}, University of North Carolina doctoral
dissertation (unpublished, 1990).

\noindent
[16]. M.R. Brown, A.C. Ottewill and D.N. Page, Phys. Rev. D{\bf 33},
2840 (1986).

\noindent
[17]. C.F.P. Lee, preprint SUSSEX-TH92/10-21 (1992).
\vfill\eject

\noindent
{\bf Appendix}

Here we outline the calculation of $\Delta S$ for a massless spin 1/2
field. We use the stress-energy tensor given in [16]. As far as we have
been able to determine, its accuracy has not been verified by an exact
numerical analysis, unlike the two cases we treated in the body of the text.
This tensor has also been used in a calculation similar to the one
presented here in [17], where qualitatively different results were
obtained for the entropy $\Delta S$.

The stress-energy tensor is given by
$$T^t_t = -{1 \over 3}aT^4_H\, ({7 \over 8}) \left(3 + 6w + 9w^2 + 12w^3
 + {135 \over 7}w^4 + {186 \over 7}w^5 - 69w^6 \right), \eqno(A1)$$

$$T^r_r = {1 \over 3}aT^4_H\, ({7 \over 8}) \left( 1 + 2w + 3w^2
-{52 \over 7}w^3 - 5w^4 - {18 \over 7}w^5 + {15 \over 7}w^6 \right),
\eqno(A2)$$

$$T^{\theta}_{\theta} = T^{\phi}_{\phi} = {1 \over 3}aT^4_H\, ({7 \over 8})
\left( 1 + 2w + 3w^2 - {45 \over 7}w^3 - {45 \over 7}w^4 +
{62 \over 7}w^5 + 23w^6 \right). \eqno(A3)$$

We find for $\mu$ and $\rho$

$$K\mu_f = {7 \over 8} \left({2 \over 3}w^{-3} + 2w^{-2} + 6w^{-1}
- 8\ln(w) - {90 \over 7}w -{62 \over 7}w^2 + {46 \over 3}w^3 - {16
\over 7} \right), \eqno(A4)$$

$$K\rho_f = {7 \over 8} \left({2 \over 3}w^{-2} + 4w^{-1}
- 8\ln(w) - {200 \over 21}w -{50 \over 7}w^2 + {52 \over 7}w^3 + {32
\over 7} \right), \eqno(A5)$$

where the subscript ``$f$" denotes ``fermion". The formulas for temperature
and inverse temperature have the same form as before with
$n_f = -4$. The quantity $\Delta S$ enjoys all the same basic properties
as for the conformal scalar and vector fields. It is given by

$$\Delta S_f = {{8\pi} \over K} ({7 \over 8})\,
\left( {8 \over 9}w^{-3} + {8 \over 3}w^{-2} + 8w^{-1} +
{24 \over 7}\ln(w) - {200 \over 21}w - {56 \over 7}w^2
+{800 \over 63}w^3 - {424 \over 63} \right), \eqno(A6)$$
and is positive.

\end{document}